\PassOptionsToPackage{numbers,sort&compress}{natbib}
\documentclass[preprint,12pt]{elsarticle}
\usepackage[total={6.25in, 8.2in}]{geometry}
\usepackage{amssymb}
\usepackage{multirow}
\usepackage{caption} 
\usepackage{amsmath} 

\usepackage{setspace}
\makeatletter
\def\ps@pprintTitle{%
   \let\@oddhead\@empty
   \let\@evenhead\@empty
   \def\@oddfoot{\hfill\thepage\hfill}
   \let\@evenfoot\@oddfoot}
\makeatother

\begin{document}

\begin{frontmatter}

\title{Percolating Corrosion Pathways of Chemically Ordered NiCr Alloys in Molten Salts}

\author[1]{Hamdy Arkoub\corref{cor1}}
\author[2]{Jia-Hong Ke\corref{cor2}}
\author[3]{Kaustubh Bawane}
\author[1]{Miaomiao Jin}
\affiliation[1]{organization={Department of Nuclear Engineering, The Pennsylvania State University},
            city={University Park},
            state={PA},
            postcode={16802},            
            country={USA}}
\affiliation[2]{organization={Computational Mechanics and Materials Department, Idaho National Laboratory},
            city={Idaho Falls},
            state={ID},
            postcode={83415},            
            country={USA}} 
\affiliation[3]{organization={Advanced Characterization Department, Idaho National Laboratory},
            city={Idaho Falls},
            state={ID},
            postcode={83415},            
            country={USA}}           
\cortext[cor1]{Corresponding author: haa5492@psu.edu}
\cortext[cor2]{Corresponding author: jiahong.ke@inl.gov}

\begin{abstract}

Recent experiments have shown that chemical ordering in NiCr alloys can significantly accelerate corrosion in molten salt environments. However, the underlying mechanisms remain poorly understood. Using reactive molecular dynamics and first-principles calculations, we show that long-range ordered Ni\textsubscript{2}Cr in Ni–33at.\%Cr alloys corrodes far more rapidly in FLiNaK salt at 800 $\mathrm{^oC}$ than short-range ordered or random solid solutions. This accelerated attack originates from percolating Cr pathways that enhance near-surface diffusion and a lowered energetic barrier for Cr dissolution, as confirmed by first-principles calculations. Contrary to earlier explanations that attributed this behavior to residual stresses, our stress-free simulations demonstrate that ordering alone accelerates the degradation. These results establish percolation as a critical link between chemical ordering and corrosion kinetics, offering a mechanistic basis for experimental observations.

\end{abstract}


\end{frontmatter}

\section{Introduction}
\label{sec:sample1}

Ni-based alloys are widely considered structural materials for high-temperature advanced energy systems due to their excellent strength, creep resistance, and tolerance to radiation-induced degradation \cite{yvon2009structural, rosenthal1972development}. Their applications in molten salt environments \cite{delpech2010molten, zohuri2020generation,williams2006assessment, delpech2010molten} at elevated temperatures exhibit critical challenges due to corrosion of structural components \cite{olson2009materials, guo2018corrosion}. Unlike aqueous environments, where corrosion may be mitigated by passivating oxide layers, molten salts destabilize these protective films, exposing the metal surface to direct attack by aggressive anions \cite{sohal2010engineering}. In fluoride salts, corrosion is typically driven by fluorine’s strong affinity for Cr, which forms soluble complexes (e.g., CrF\textsubscript{2}, CrF\textsubscript{3}), leading to selective Cr dissolution, dealloying, and eventual mechanical and microstructural degradation \cite{olson2009materials, yin2018first, chan2022insights,zheng2015corrosion, chan2024morphological}.

Atomic ordering, whether short-range order (SRO) or long-range order (LRO), is an intrinsic outcome of alloy thermodynamics and can strongly influence material properties. In Ni–Cr alloys, LRO phases such as the MoPt\textsubscript{2}-type Immm Ni\textsubscript{2}Cr form under specific thermal aging conditions \cite{kolotushkin2003effect, sundararaman1999precipitation}, and have been extensively studied for their effects on mechanical behavior, including ductility and hardness \cite{xie2008precipitation, young2016effect, vo2021deformation}. SRO, though more subtle, has also been shown to alter macroscopic properties related to phase stability and corrosion susceptibility, for instance promoting intergranular stress corrosion cracking (SCC) in Alloy 600 \cite{kim2015effect} and modifying the thermophysical behavior of Ni–Cr–Mo alloys \cite{zhilyakov2020relation}. These findings suggest that ordering at multiple length scales can shape corrosion-related responses.

Despite this, the effect of ordering on corrosion remains poorly understood and in some cases contradictory. In Ni–Cr–Mo alloys, the formation of Ni\textsubscript{2}(Cr,Mo)-type LRO phases has been linked to improved corrosion resistance in chloride-based molten salts and aqueous systems \cite{polovov2019effect, tawancy2018correlation}. By contrast, Teng et al. \cite{teng2023accelerated} reported that Ni\textsubscript{2}Cr LRO phases accelerate corrosion in binary Ni–Cr alloys exposed to NaCl–MgCl\textsubscript{2}, and Young et al. \cite{young2016effect} found that LRO increases susceptibility to SCC by three orders of magnitude relative to disordered alloys. These discrepancies raise a fundamental question: is corrosion resistance governed by the intrinsic effect of LRO, or by alloying additions such as Mo? Proposed explanations, including anisotropic corrosion behavior and internal strain associated with ordering, remain speculative. To date, no systematic studies have addressed how SRO and LRO influence corrosion in high-temperature molten salt environments. 

Given that ordering can alter diffusion, phase stability, and corrosion susceptibility, clarifying its role in molten salt corrosion is essential for both fundamental understanding and alloy design. Atomistic modeling provides a means to directly probe molten salt corrosion mechanisms. While density functional theory (DFT) has provided valuable insights into early-stage corrosion and adsorption energetics, its computational cost limits its applicability to small systems and short time scales \cite{yin2018first,startt2021ab, schneider2024mechanistic}. Reactive molecular dynamics (RMD) with ReaxFF force fields extends these capabilities to chemically reactive, large-scale simulations that capture interfacial reactions and atomic transport over nanoseconds \cite{van2001reaxff, senftle2016reaxff}. Previous ReaxFF studies have established that Ni–Cr corrosion in molten FLiNaK is driven primarily by strong Cr–F bonding, which accelerates Cr dissolution \cite{arkoub2024reactive}. These processes are largely surface-controlled, governed by near-surface atom transport \cite{arkoub2025surface}.

While those existing reactive simulations have been successfully applied to binary systems including Ni–Cr, Fe–Ni, Fe–Cr, and Ni–Al, they have so far assumed only randomly distributed alloy configurations \cite{arkoub2025surface, zhu2025atomic, cheng2025effect, qu2025atomic, ai2021oxidation, chen2021effect}, without considering the influence of chemical ordering. In this work, we fill this gap by applying ReaxFF-based RMD to compare disordered, SRO, and LRO Ni–33at.\%Cr alloys in molten FLiNaK. The results show that LRO Ni\textsubscript{2}Cr undergoes markedly accelerated Cr dissolution compared to SRO and random solid solutions (RSS). It is found that this behavior arises from percolating Cr pathways that enhance the transport of subsurface Cr to the metal-salt interface. Complementary first-principles calculations confirm that LRO also lowers the energetic barrier for Cr dissolution. Together, those results reveal percolation pathways as the critical link between chemical ordering and accelerated molten salt corrosion, providing a mechanistic explanation for experimental observations.

\section{Methods}
 \subsection{RMD simulations}
\label{sect:MD}

RMD simulations were performed using the Large-scale Atomic/Molecular Massively Parallel Simulator (LAMMPS) code \cite{thompson2022lammps} with the ReaxFF force field \cite{van2001reaxff, senftle2016reaxff}. We employed a ReaxFF potential developed and validated for the NiCr–FLiNaK system \cite{arkoub2024reactive}, which can accurately reproduce the thermophysical properties of the salt and the corrosion behavior of NiCr alloys in molten fluoride environments \cite{arkoub2024reactive, arkoub2025surface}. To probe the role of chemical ordering, Ni–33 at.\% Cr slabs with a (110) surface orientation were constructed. The (110) surface was chosen because our previous study identified it as more susceptible to corrosion than the (100) and (111) orientations \cite{arkoub2025surface}, making it the most suitable for capturing dissolution within accessible MD timescales. Alloy configurations representing RSS, SRO, and LRO Ni\textsubscript{2}Cr were generated. SRO and LRO structures were created using Monte Carlo simulations (see Sec. \ref{sec:MC}), containing 1,134 Cr and 2,268 Ni atoms. 

\begin{figure}[!ht]
	\centering
	\includegraphics[width=0.9\textwidth]{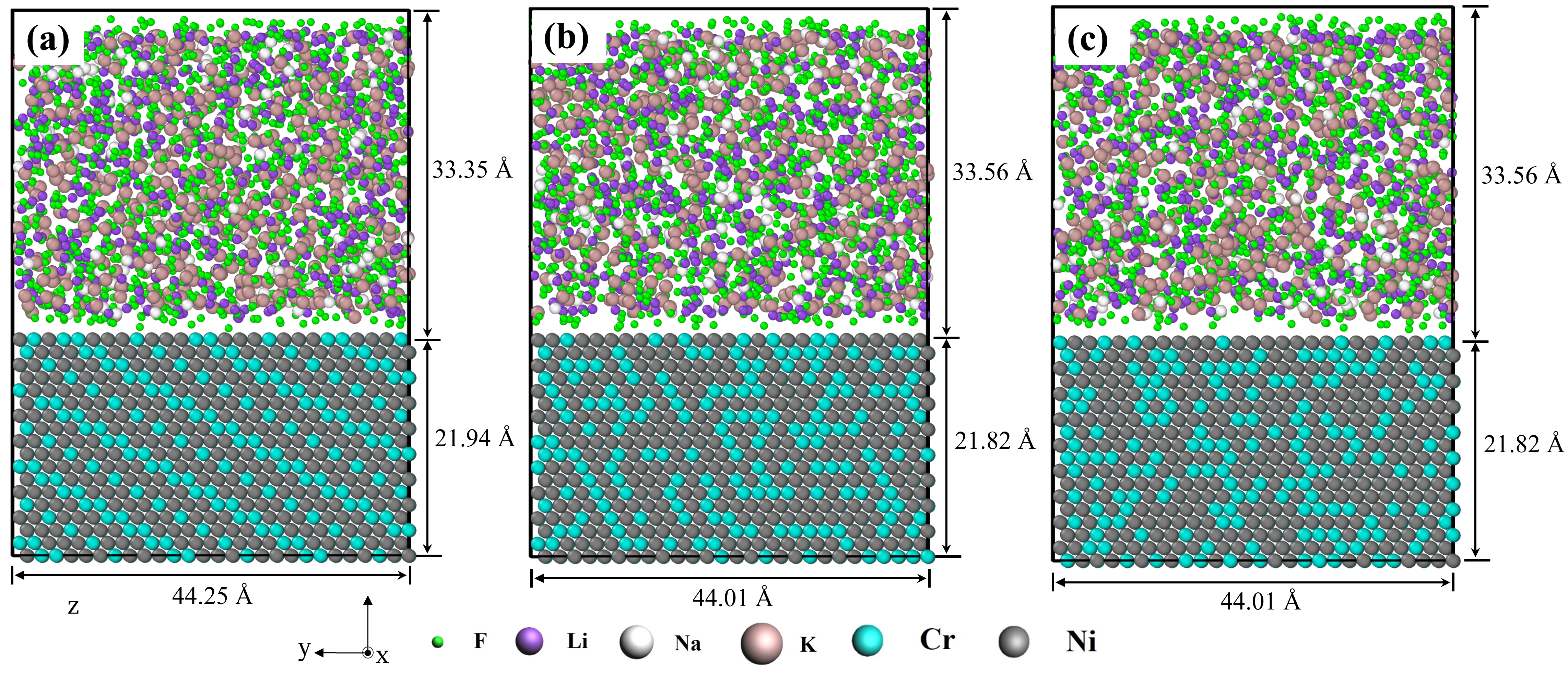}
	\caption{Optimized simulation cells for (a) LRO, (b) SRO, and (c) RSS (110)-surface-oriented Ni–Cr alloys in contact with FLiNaK salt at 800 \textsuperscript{o}C and atmospheric pressure.}
	\label{fig:model}
\end{figure}

Following the construction of the alloy with slab geometry, equilibration was performed for 50 ps at 800\textsuperscript{o}C and atmospheric pressure using the NPT ensemble with a time step of 0.25 fs and periodic boundary conditions \cite{berendsen1984molecular, nose1984unified}. The relaxed cell dimensions were slightly larger for LRO (44.27 $\times$ 40.25 $\times$ 23.24 {\AA}) than for SRO and RSS (44.09 $\times$ 40.08 $\times$ 23.14 {\AA}), due to local relaxation in ordered structures \cite{gwalani2016experimental}. Molten FLiNaK was generated using the Packmol package \cite{martinez2009packmol} in a typical ratio of 46.2–11.5–42 mol\% LiF, KF, and NaF \cite{williams2006assessment}, corresponding to 558 LiF, 504 KF, and 138 NaF molecules in the system. The salt was equilibrated separately at 800\textsuperscript{o}C and atmospheric pressure with a 0.25 fs timestep using the Berendsen barostat \cite{berendsen1984molecular} and Nose–Hoover thermostat \cite{nose1984unified}. During equilibration, the $x$ and $y$ dimensions were constrained to match the alloy slab. After equilibration, the molten salt was placed atop the alloy surfaces to construct the alloy–salt simulation cells, as shown in Figure \ref{fig:model}.

Corrosion simulations were carried out under canonical settings at 800\textsuperscript{o}C, using the Nose–Hoover thermostat \cite{nose1984unified}. This temperature is relevant to the typical operating conditions of molten salt reactors, and also allows for observation of corrosion behavior within nanosecond RMD timescales \cite{wright2018status, serp2014molten, arkoub2025surface}. Although Ni\textsubscript{2}Cr is thermodynamically stable below 590 °C \cite{nash1986cr}, prior experiments show it persists metastably at higher temperatures before decomposing at 900 °C \cite{miller2018phase}. To approximate bulk behavior and prevent artificial slab deformation, the bottom two alloy layers were fixed during the simulations. Fluoride surface coverage was quantified using cutoff distances equal to 80\% of the sum of van der Waals radii: 2.776 {\AA} for Cr–F and 2.48 {\AA} for Ni–F. A metal atom was considered dissolved if, based on its bonding environment with fluorine and local coordination, it had fewer than two neighboring metal atoms and formed stable bonds with fluorine. For statistical reliability, ten independent simulations were performed for each alloy configuration. Visualization and analysis were performed using the OVITO software package \cite{stukowski2009visualization}.

\subsection{Chemical Order Generation}
\label{sec:MC}

SRO and LRO configurations of Ni–33at.\%Cr were generated using canonical Monte Carlo (MC) simulations with the Metropolis algorithm. Simulations began from a random solid solution in a supercell. The supercell was constructed from the face-centered cubic (FCC) lattice with a (110) orientation using the transformation matrix {[-7, 9, -10], [-7, 0, 20], [21, 0, 0]}, making it suitable for subsequent slab construction in Sec. \ref{sect:MD}. 

During each MC step, a Ni–Cr pair was randomly chosen for a trial swap. If the swap increased the formation energy by $\Delta E$, it was accepted following the probability $\exp(-\Delta E/k_B T)$, where $k_B$ is the Boltzmann constant; otherwise, the swap was rejected. Formation energies were calculated on-the-fly using a DFT-informed cluster expansion (CE) Hamiltonian constructed for the Ni–Cr binary system \cite{ke2023microstructure,barnard2014atomistic}. The comparison between CE and DFT formation energies, shown in Supplementary Materials (SM) Figure S1(a), confirms the accuracy of the CE model.

For the canonical MC simulations, we followed the method and equilibration conditions used in \cite{ke2025role,ke2023effects}. The ensemble averages of the formation energy were measured every $N$ Monte Carlo steps, where $N$ is the number of sites in the supercell with variable occupation. After a sufficient number of Monte Carlo steps and iterations, the ensemble averages of the formation energy converged, achieving equilibrium configuration. The simulations were stopped once the convergence of the formation energy per unit cell reached $3 \times 10^{-5}$ eV/atom. The resulting SRO configuration was validated by comparing the calculated Warren–Cowley SRO parameters with experimental values obtained from in situ neutron scattering measurements of Ni–33 at.\% Cr at 650\textdegree C \cite{caudron1992situ}, as shown in SI Figure S1(b). Our MC simulations accurately captured the first three shells of the Warren-Cowley SRO parameters \cite{cowley1965short}. While the fourth-shell parameter was more negative than experimental values, the overall agreement supported the use of the derived SRO configuration to distinguish it from a fully RSS. The LRO configuration was derived by the same canonical MC method and convergence condition at 550 \textdegree C, which is below the order-disorder transition temperature \cite{miller2018phase}. The resulting LRO structure (the alloy matrix in Figure \ref{fig:model} (a)) demonstrates Pt$_2$Mo type ordering \cite{turchi2006modeling}. Those alloy structures were then used for constructing the final supercells as described in Sec. \ref{sect:MD}.

\subsection{First-principles calculations}

DFT calculations were performed using the Vienna \textit{Ab initio} Simulation Package (VASP) \cite{kresse1993ab,kresse1996efficient} with projector augmented-wave (PAW) potentials and a plane-wave energy cutoff of 400 eV. Spin-polarization was applied to all calculations, and exchange correlation interactions were treated within the generalized gradient approximation (GGA) using the Perdew–Burke–Ernzerhof (PBE) functional \cite{perdew1996generalized}. Geometry optimizations employed the conjugate-gradient method with energy and force convergence criteria of $10^{-6}$ eV and 0.01 eV/Å, respectively. Both the LRO and disordered Ni–33 at.\% Cr (100) slabs contained four atomic layers along the $z$-direction, giving a total of 36 atoms per supercell. The disordered alloy was represented by a Special Quasirandom Structure (SQS), following the method by Zunger and Wei \cite{zunger1990special, wei1990electronic}. The SQS was generated via MC swapping to reproduce random alloy pair correlations up to the fourth-nearest neighbor shell, with deviations below 0.02 from ideal randomness.

The FLiNaK salt model (14 LiF, 4 NaF, 13 KF) was first equilibrated by \textit{Ab initio} molecular dynamics (AIMD) at 700 °C for 0.5 ps using a 0.5 fs timestep under the NVT ensemble. The equilibrated salt was then placed atop the alloy slab, forming a supercell containing 99 atoms. The final supercell dimensions were 7.55 × 7.55 × 29.00 Å, and periodic boundary conditions were applied in all directions. The dissolution process of a Cr adatom in contact with molten FLiNaK was evaluated for both LRO and SQS using the constrained stepwise approach, following our previous work \cite{arkoub2024first}. Namely, during dissolution simulations, a Cr adatom was incrementally displaced along the surface normal in 0.5 Å steps, with all other atoms energy minimized at each stage. Brillouin-zone sampling employed a $\Gamma$-centered $3\times3\times1$ k-point mesh. To maintain a consistent interfacial environment, the same equilibrated salt configuration was used for both the LRO and SQS slabs.

\section{Results and Discussion}

\begin{figure}[!ht]
	\centering
	\includegraphics[width=0.52\textwidth]{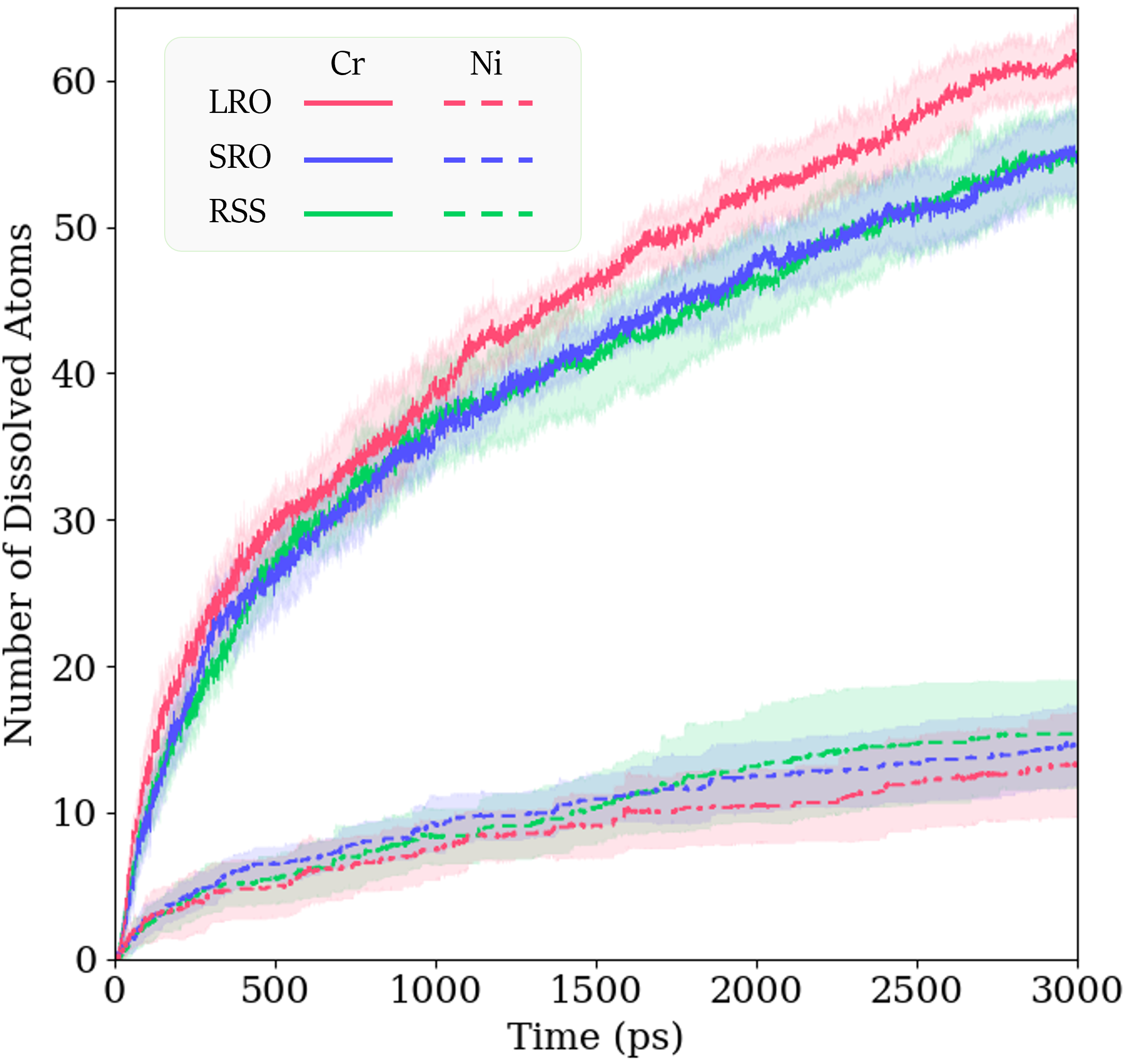}
	\caption{Time evolution of Cr and Ni dissolution from the NiCr alloys with LRO, SRO, and RSS configurations, in molten FLiNaK at 800\textsuperscript{o}C. Shaded regions indicate standard deviations.}
	\label{fig:dissolution}
\end{figure}

Figure \ref{fig:dissolution} shows the time evolution of Cr and Ni dissolution from Ni–33at.\%Cr alloys with LRO, SRO, and RSS configurations in molten FLiNaK at 800 \textsuperscript{o}C. During the first nanosecond, all three alloys exhibit similar short-time dissolution, though the LRO system shows a slightly higher tendency for Cr loss. By 2 ns, a clear trend emerges: the LRO alloy dissolves noticeably more Cr than SRO and RSS, which display similar behavior within statistical uncertainty. By 3 ns, the LRO configuration exhibits substantially greater Cr dissolution, whereas the SRO and RSS alloys remain nearly the same. Ni dissolution remains minimal across all configurations, consistent with prior computational and experimental studies \cite{fayfar2023situ, mills2024elucidating, arkoub2024reactive, arkoub2025surface}.

The comparable behavior of SRO and RSS alloys suggests that SRO does not strongly influence early-stage corrosion under the nanosecond timescales of RMD. At this stage, corrosion proceeds primarily through local alloy–salt interactions SRO and RSS alloys, before the structural ordering effects begin to emerge. In contrast, the pronounced corrosion of LRO alloys reproduces the experimentally observed susceptibility of Ni\textsubscript{2}Cr to accelerated degradation in molten salt \cite{teng2023accelerated}. This behavior was previously attributed to residual stresses from precipitate formation \cite{teng2023accelerated}, yet the current simulations capture this feature without introducing stress. It indicates that the atomic arrangement in the ordered phase itself in the molten salt environment drives rapid dealloying without extra external driving force. It should be noted that, LRO does not always weaken corrosion resistance, for example, in Ni–Cr–Mo alloys, ordered phases have been reported to improve durability \cite{polovov2019effect, tawancy2018correlation}. Hence, the effect of ordering on corrosion is system-specific.

\begin{figure}[!ht]
	\centering
	\includegraphics[width=0.9\textwidth]{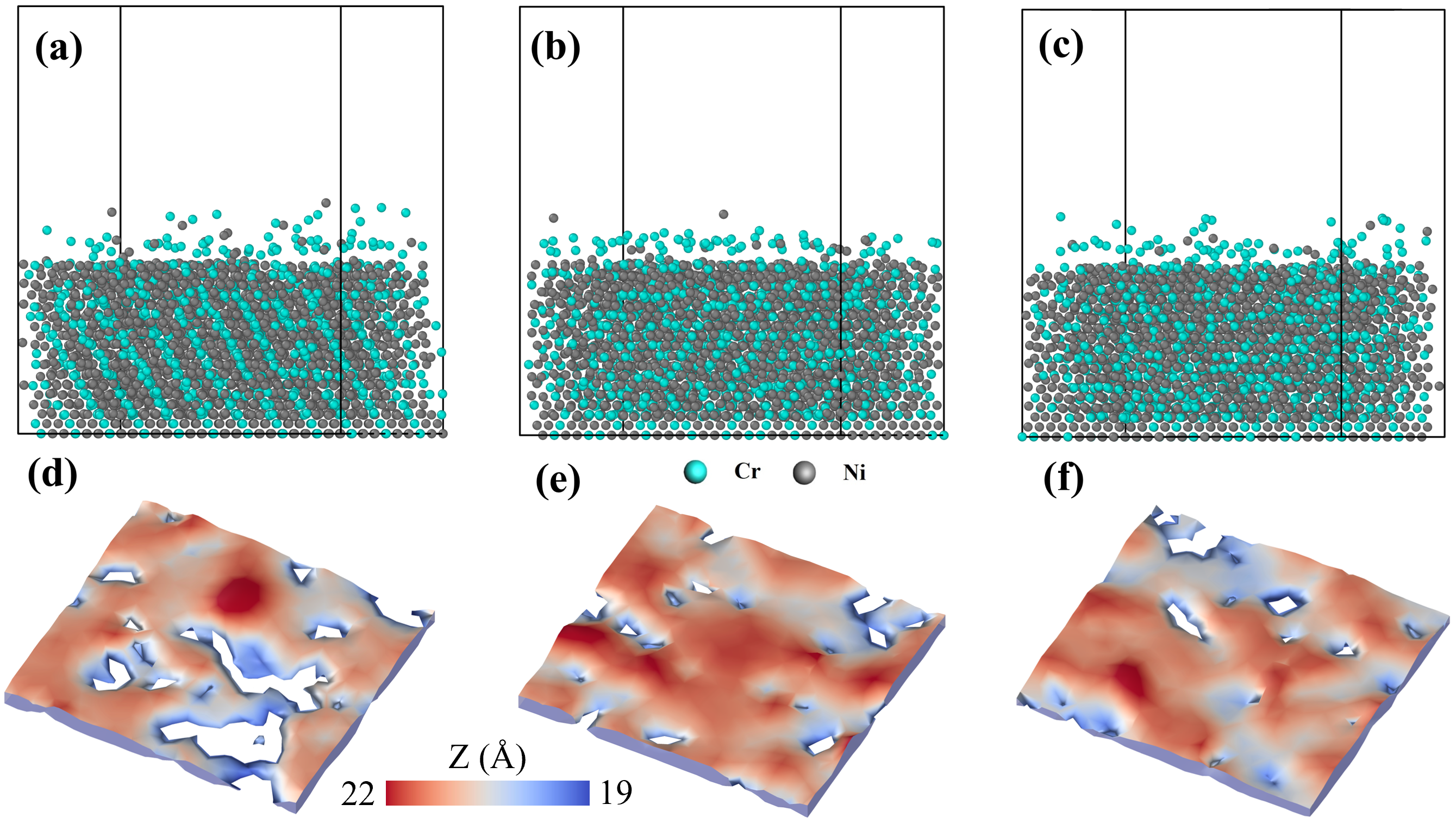}
	\caption{Snapshots of NiCr slabs with (a) LRO, (b) SRO, and (c) RSS after 3 ns at 800°C in contact with molten FLiNaK salt. Salt ions are removed for clarity. (d-f) Corresponding surface meshes of the LRO, SRO, and RSS slabs, respectively, color-coded by the local $Z$ coordinates. Each surface morphology is reconstructed by isolating the uppermost 3.5 Å layer of each slab.}
	\label{fig:surface}
\end{figure}

Figure \ref{fig:surface} shows snapshots and surface meshes of the LRO, SRO, and RSS slabs after 3 ns of exposure to molten salt. In the LRO alloy (Figure \ref{fig:surface}a), LRO is disrupted within the surface layers directly contacting the salt due to interfacial reactions and preferential Cr dissolution, which promote enhanced diffusion and disordering near the interface. The interior bulk, however, largely preserves its ordered structure with minimal disruption. Surface mesh visualizations (Figures \ref{fig:surface}d–f) highlight the morphological consequences: the LRO slab develops a rough, porous surface with deep pits, while the SRO and RSS slabs remain comparatively smooth. This enhanced degradation corresponds directly to the higher Cr dissolution seen in Figure \ref{fig:dissolution}. As shown previously \cite{arkoub2025surface}, Cr leaching generates terrace vacancies that rapidly migrate and coalesce into pits. This pitting morphology, rather than uniform planar dissolution, arises from the competition between dissolution kinetics and surface diffusion. Preferential Cr removal produces localized vacancies, and their rapid lateral migration, facilitated by surface diffusion and the slower dissolution of the Ni-rich matrix, drives pit growth and deepening. As vacancies coalesce and redistribute, the surface evolves into a porous network. In the LRO alloy, this process is accelerated, producing a porous surface network. These observations agree with experimental findings by Teng et al. \cite{teng2023accelerated}, where TEM analysis revealed selective disruption of LRO regions via Cr leaching, and are also consistent with prior dealloying studies showing that preferential Cr dissolution yields Ni-rich porous structures \cite{mccue2016dealloying, liu2021formation, liu2023temperature}.

\begin{figure}[!ht]
	\centering
	\includegraphics[width=0.9\textwidth]{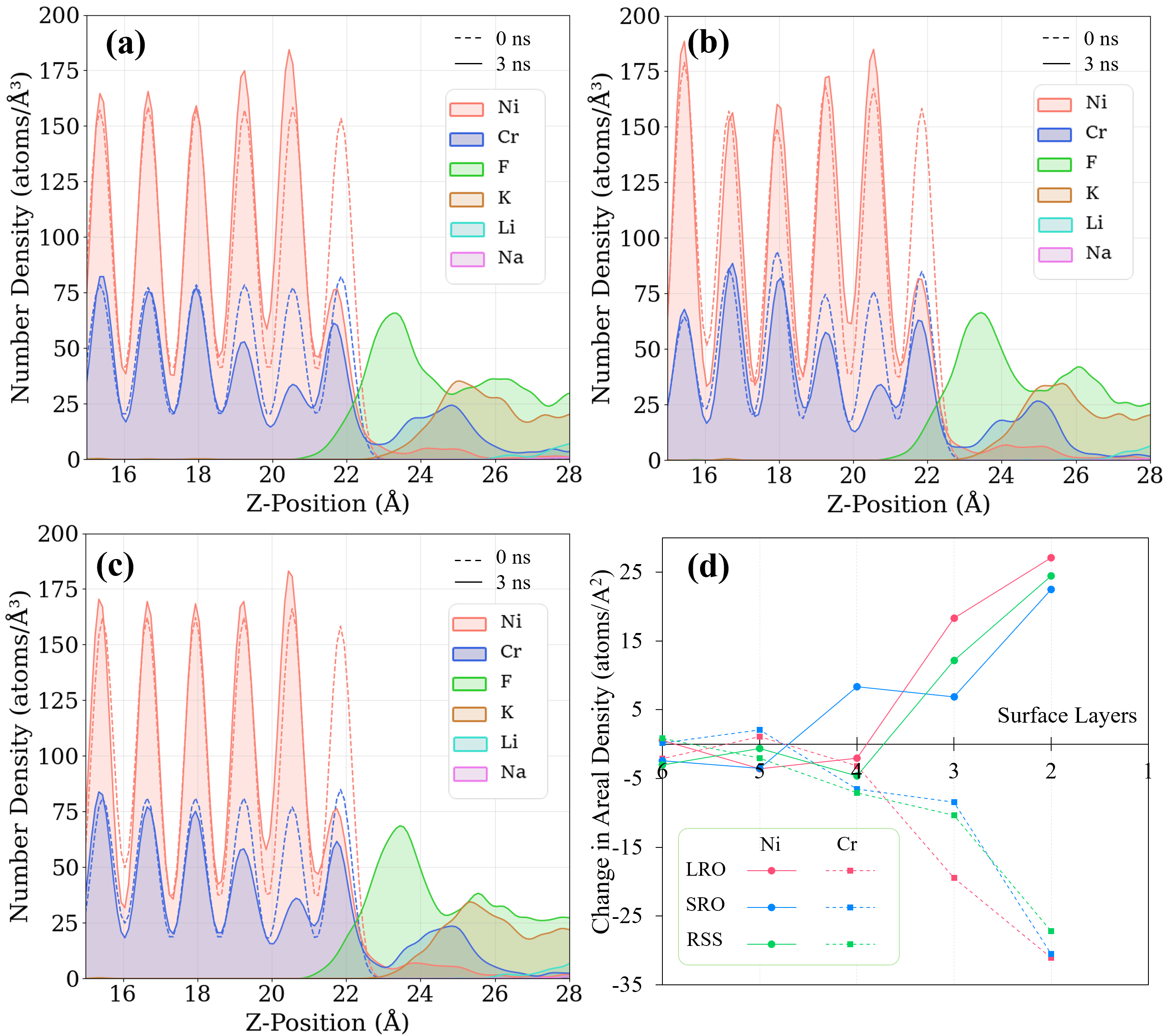}
	\caption{Average atomic number density profiles along the $Z$-direction of NiCr alloys at 800 $\mathrm{^o}$C with (a) LRO, (b) SRO, and (c) RSS configurations at 0 ns (dahsed) and 3 ns (solid). (d) Change in areal density of Ni and Cr across the toplayers after 3 ns.}
	\label{fig:atomic}
\end{figure}

To examine corrosion-induced compositional changes, atomic number density profiles along the slab normal are shown in Figure \ref{fig:atomic}. Initially (dashed lines), all slabs exhibit well-defined Ni and Cr peaks near the surface. However, after 3 ns (solid lines), the Cr peak intensity decreases while Ni intensity slightly increases, indicating interlayer atomic rearrangement. Cr is depleted from the first few surface layers due to its outward diffusion to the salt interface. Upon closer inspection, occasional subsurface vacancies are also observed beneath the topmost layers, which could be precursors to the experimentally observed voids in sub-surface and grain-boundary regions \cite{yang2023one,wang2022integrated,bawane2022visualizing,ghaznavi2022alloying}. Representative snapshots illustrating these subsurface vacancies are provided in the SM Figure S2. The degree of atom redistribution depends strongly on ordering: Figure \ref{fig:atomic}(d) shows that Cr loss and Ni gain are much greater in the LRO slab than in SRO or RSS, which suggests that LRO alloys undergo faster surface diffusion. The net decrease in the combined Ni and Cr number density corresponds to the atoms that have dissolved into the salt. The first peak in the salt distribution (Figure \ref{fig:atomic}(a–c)) arises from F ions adsorbed onto the surface through interactions with Cr. This behavior is consistent with previous studies \cite{arkoub2024reactive, arkoub2025surface,wang2025exploring, yin2018first}. Notably, the fluorine adsorption peaks (magnitude and position) are unaffected by the slab’s chemical ordering.

\begin{figure}[!ht]
	\centering
	\includegraphics[width=1.0\textwidth]{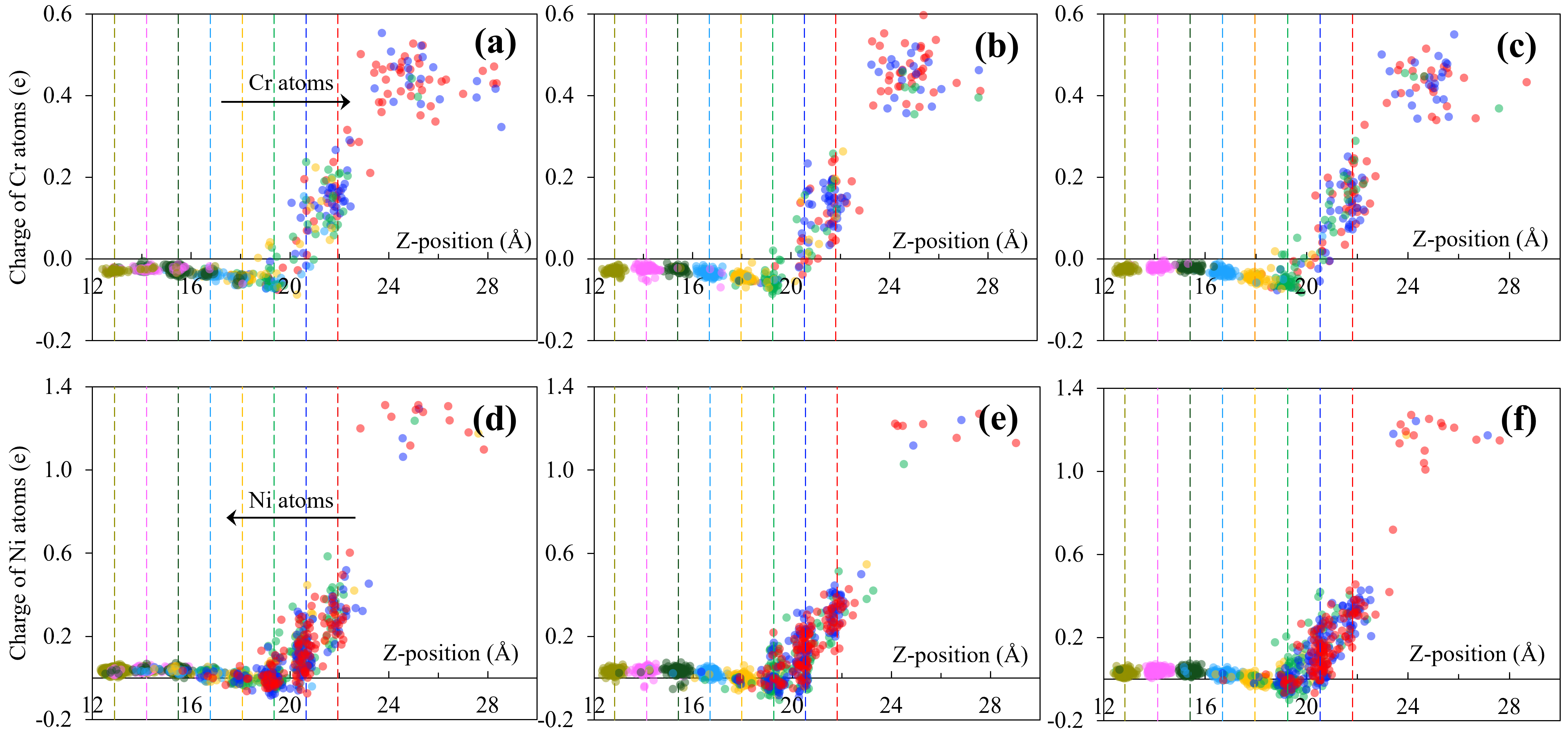}
	\caption{Charge distribution along the $Z$-direction of the NiCr alloys after 3 ns at 800 $\mathrm{^o}$C: (a,d) LRO, (b,e) SRO, and (c,f) RSS slabs. (a-c) Cr charges and (d-f) Ni charges. Dashed lines mark the initial layer positions. Atom charges are color-coded based on the original layers in the initial perfect structure.}
	\label{fig:charge}
\end{figure}

The impact of atomic ordering on corrosion-induced elemental redistribution is further illustrated by the charge distribution profiles of Cr and Ni atoms, shown in Figure \ref{fig:charge}. In these plots, atomic charges are color-coded based on the atoms' original layer in the pristine structure, with vertical dashed lines marking the initial layer positions. This visualization shows the transport of atoms across layers as corrosion progresses. For all cases, atoms charges in the bulk remain close to zero, while significant charge perturbations appear at the interface. Consistent with prior studies \cite{arkoub2024first,arkoub2024reactive}, dissolved Ni carries a higher residual positive charge (+1.3e) than Cr (+0.4e), reflecting weaker fluoride complexation for Ni compared to the more delocalized charge state of Cr in CrF\textsubscript{x} species. Also, across all alloy types, near-surface Cr atoms migrate outward toward the salt interface, while Ni atoms diffuse inward, while the rates of those physical processes differ noticeably between systems. In the LRO case (Figure \ref{fig:charge}(a,d)), the pronounced cross-layer atomic transport leads to enhanced Cr dissolution and smearing of the surface structures. By contrast, the SRO and RSS alloys (Figures \ref{fig:charge}b,c,e,f) exhibit more confined transport, with most atoms remaining close to their initial layers. For example, Cr atoms originally in the third layer of the LRO alloy (green dots) are nearly depleted, while the same layer in SRO and RSS retains significant Cr. These results provide direct evidence that LRO facilitates fast Cr diffusion pathways toward the salt, accelerating Cr dissolution.

To gain deeper insight into the atomic-scale mechanisms underlying the relative rates of near-surface processes, we next quantify near-surface mobility using mean square displacement (MSD) analysis. Specifically, the MSD of Ni and Cr atoms in the top three surface layers (excluding those that dissolved into the salt) was calculated. The corresponding diffusion coefficients ($D$) were obtained by linear regression of the MSD as a function of time ($t$):

\[\mathrm{MSD}=\frac{1}{N}\langle \sum_{i}^{N} (r_i(t)-r_i(0))^2 \rangle\qquad(1)\]
\[\mathrm{D}=\frac{1}{2d} \cdot \frac{d}{dt} \langle (r_i(t)-r_i(0))^2 \rangle\qquad(2)\]
where $N$ is the number of atoms considered in MSD statistics, $d$ is the number of dimensions (=3), and $r_{i}(t)$ is the atom position at time $t$. 

\begin{figure}[!ht]
	\centering
	\includegraphics[width=1.0\textwidth]{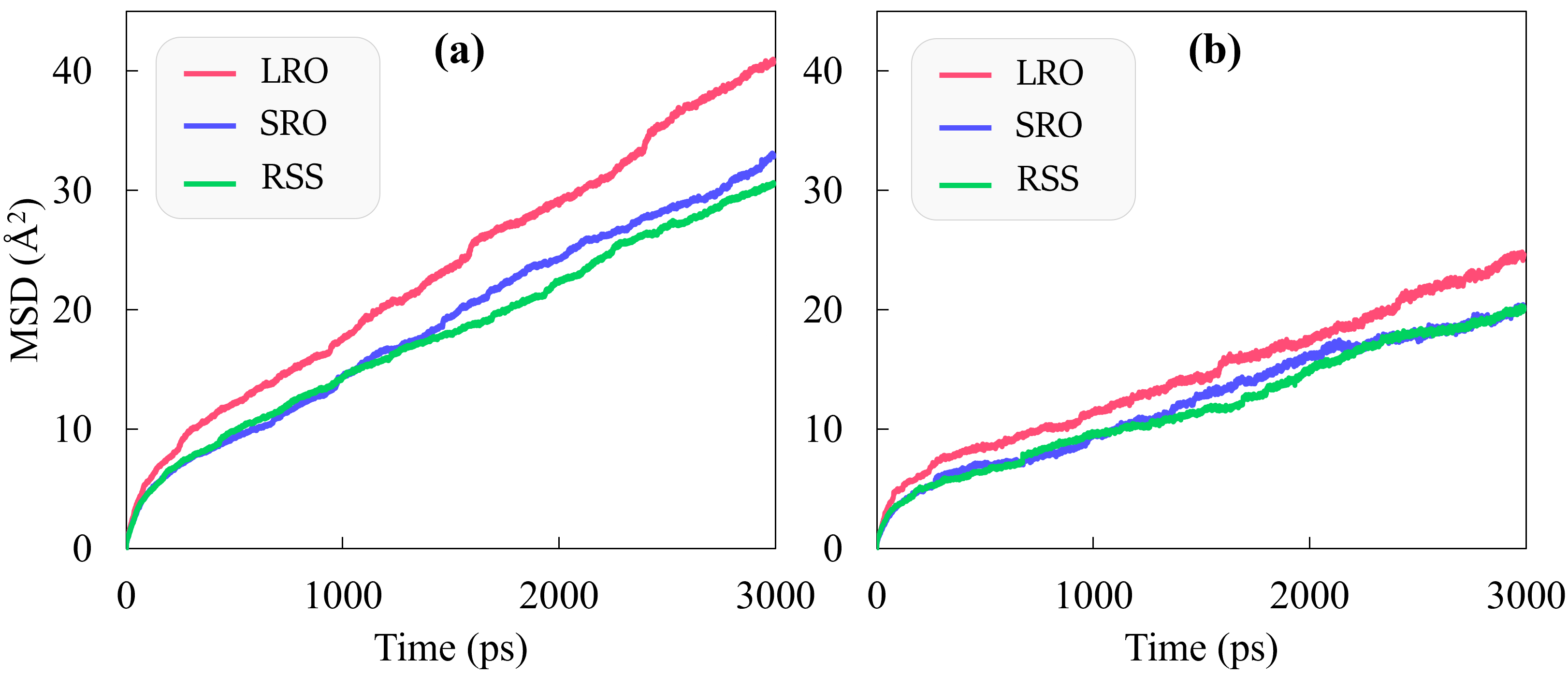}
	\caption{Average mean square displacement (MSD) of metal atoms in the top three surface layers of NiCr alloys exposed to molten FLiNaK. (a) MSD of Ni atoms; (b) MSD of Cr atoms.}
	\label{fig:msd}
\end{figure}

The results are shown in Figure \ref{fig:msd}, and the corresponding $D$ values are summarized in Table \ref{table_1}. The MSD profiles reveal that LRO alloys exhibit the highest atomic mobility for both Ni and Cr, while the SRO and RSS alloys show comparable behavior. This demonstrates that long-range ordering significantly enhances near-surface diffusion. This enhancement arises from two possible reasons: i) intrinsically higher diffusion in the ordered structure and ii) the increased Cr dissolution in LRO, which generates additional surface vacancies that promote atomic transport. From Figure \ref{fig:msd}, the influence of LRO is more pronounced for Ni diffusion than for Cr. It can also be seen that Ni exhibits a higher diffusion coefficient than Cr across all configurations, despite Cr being the faster bulk diffuser in Ni–Cr alloys \cite{tucker2010ab}. This discrepancy arises because dissolved atoms are excluded from the MSD calculations; since Cr is the primary dissolving species, its diffusivity is consequently underestimated.

\begin{table}[!ht]
\footnotesize
    \centering
    \caption{Comparison of near-surface Ni and Cr atoms Diffusion Coefficients for the three Ni-33at.\%Cr alloys.}
    \begin{tabular}{l|ll}
    \hline
    \multirow{2}{*}{Ni-33at.\%Cr Alloy} & \multicolumn{2}{l}{D (cm$^2$/s)} \\ \cline{2-3} 
                                        & Ni             & Cr           \\ \hline
    LRO                                 & 1.95 x 10$^{-7}$    & 1.08 x 10$^{-7}$  \\ 
    SRO                                 & 1.57 x 10$^{-7}$   & 9.50 x 10$^{-8}$   \\ 
    RSS                              & 1.40 x 10$^{-7}$     & 9.33 x 10$^{-8}$  \\ \hline
    \end{tabular}
    \label{table_1}
\end{table}

Although the SRO and RSS alloys exhibit nearly identical short-term dissolution behavior (Figure \ref{fig:dissolution}), MSD analysis reveals an enhancement in diffusivity for the SRO configuration, particularly when deeper layers are considered (SM Figure S3). This observation contrasts with prior studies on multi-principal element alloys (MPEAs), where SRO was generally shown to suppress bulk atomic transport by increasing local chemical heterogeneity and disrupting percolating diffusion pathways \cite{zhao2021role,zhao2019diffusion}. Compared to MPEAs, the Ni–Cr alloy exhibits relatively weak enthalpic interactions and a narrow range of site energies, leading to reduced heterogeneity in the potential energy landscape \cite{rahaman2014first,dudova2009short,fernandez2017short}. Moreover, under molten salt exposure, corrosion-induced vacancies promote vacancy-mediated diffusion over preferential pathways associated with SRO domains. This enhanced diffusivity is expected to influence long-term corrosion kinetics under extended salt exposure, where vacancy generation and redistribution dominate mass transport; such effects remain beyond the nanosecond timescales captured here.

\begin{figure}[!ht]
	\centering
	\includegraphics[width=1.0\textwidth]{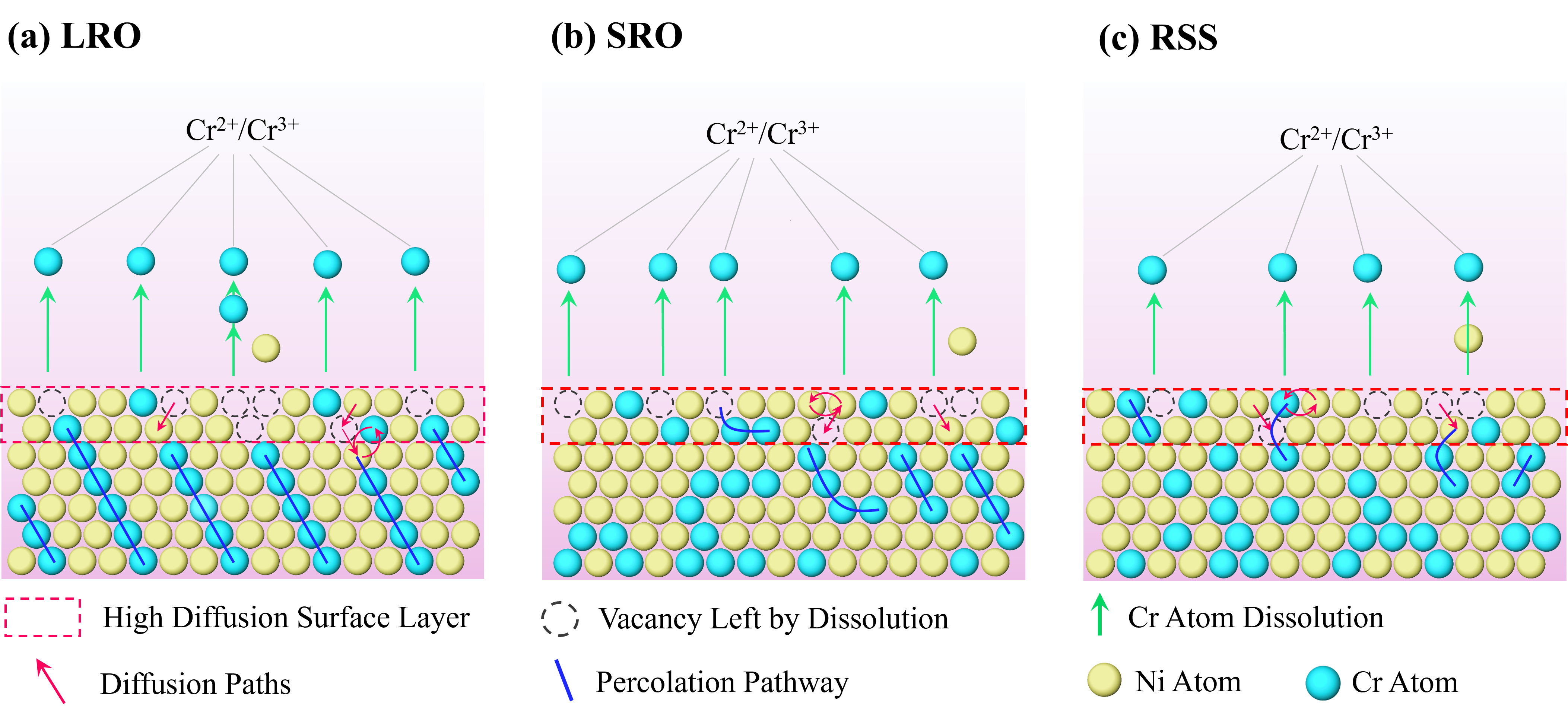}
	\caption{Schematic illustration of percolation pathways for Cr dissolution. (a) LRO alloys form continuous Cr networks that facilitate transport to the surface, while (b) SRO and (c) RSS alloys exhibit more fragmented, less connected pathways.}
	\label{fig:graph}
\end{figure}

Next, we discuss the mechanistic origin of the corrosion behavior observed in the different configurations (schematically illustrated in Figure \ref{fig:graph}). In the LRO alloy, corrosion induces pronounced cross-layer mixing in the surface region (Figure \ref{fig:charge}(a,d)), while the underlying bulk retains its ordered structure (Figure \ref{fig:surface}(a)), consistent with experiments \cite{teng2023accelerated}. Analysis of the atomic evolution reveals that this behavior follows a percolation-driven process. In the ordered Ni\textsubscript{2}Cr lattice, Cr atoms form a continuous network that connects subsurface regions to the salt interface. As corrosion proceeds, preferential Cr dissolution generates vacancies at the interface, enabling the percolating Cr network to act as a “highway” that continuously feeds Cr atoms to the interface for dissolution. This allows rapid corrosion kinetics, while the deeper layers beneath the corroded surface remain ordered. From an atomistic perspective, this percolation-driven mechanism also explains the observed diffusivity trends. The continuous Cr sublattice in LRO not only enables fast migration of Cr atoms but also indirectly enhances Ni mobility through vacancy-assisted diffusion (Figure \ref{fig:msd}(a)). In contrast, SRO and RSS alloys possess more fragmented Cr distributions that limit atomic connectivity and long-range transport. The SRO configuration, containing partially connected Cr clusters, is expected to exhibit intermediate behavior, more corrosion than RSS but less than LRO. On the nanosecond timescales accessible to RMD simulations, differences between SRO and RSS are minimal because corrosion remains localized to the topmost layers.

The mechanistic picture revealed here aligns conceptually with the percolation framework described by Xie et al. \cite{xie2021percolation}, but with a different outcome depending on the corrosion environment. In their aqueous systems \cite{xie2021percolation}, continuous Cr-rich networks enable rapid nucleation of protective oxide films, thereby enhancing corrosion resistance. In the current fluoride molten salts, where oxide passivation is thermodynamically unstable \cite{sohal2010engineering}, these same Cr networks facilitate Cr transport toward the interface, accelerating dissolution and dealloying. This percolation-driven dealloying process, involving the surface and inward diffusion of Ni, is the key mechanism for forming bicontinuous structures in molten salts, as demonstrated by Liu et al. \cite{liu2021formation}. Thus, the presence of percolating pathways likely universally governs transport kinetics in different alloys with LRO. In ternary Ni–Cr–Mo alloys, Ni\textsubscript{2}(Cr,Mo)-type ordering was found to improve corrosion resistance \cite{polovov2019effect, tawancy2018correlation}. This could be explained that Mo modifies the Cr network, which suppresses Cr percolation and eventually resists Cr dissolution. Recent experimental work by Liu et al. \cite{liu2025enhancing} further supports this interpretation: Mo enrichment in stainless steels was shown to trap Cr, which suppresses Cr outward diffusion and corrosion in molten salts. By contrast, in the binary Ni–Cr system, Ni\textsubscript{2}Cr ordering produces an uninterrupted Cr lattice, lowering the energetic barrier for Cr migration and thereby accelerating dissolution. 

\begin{figure}[!ht]
	\centering
	\includegraphics[width=1.0\textwidth]{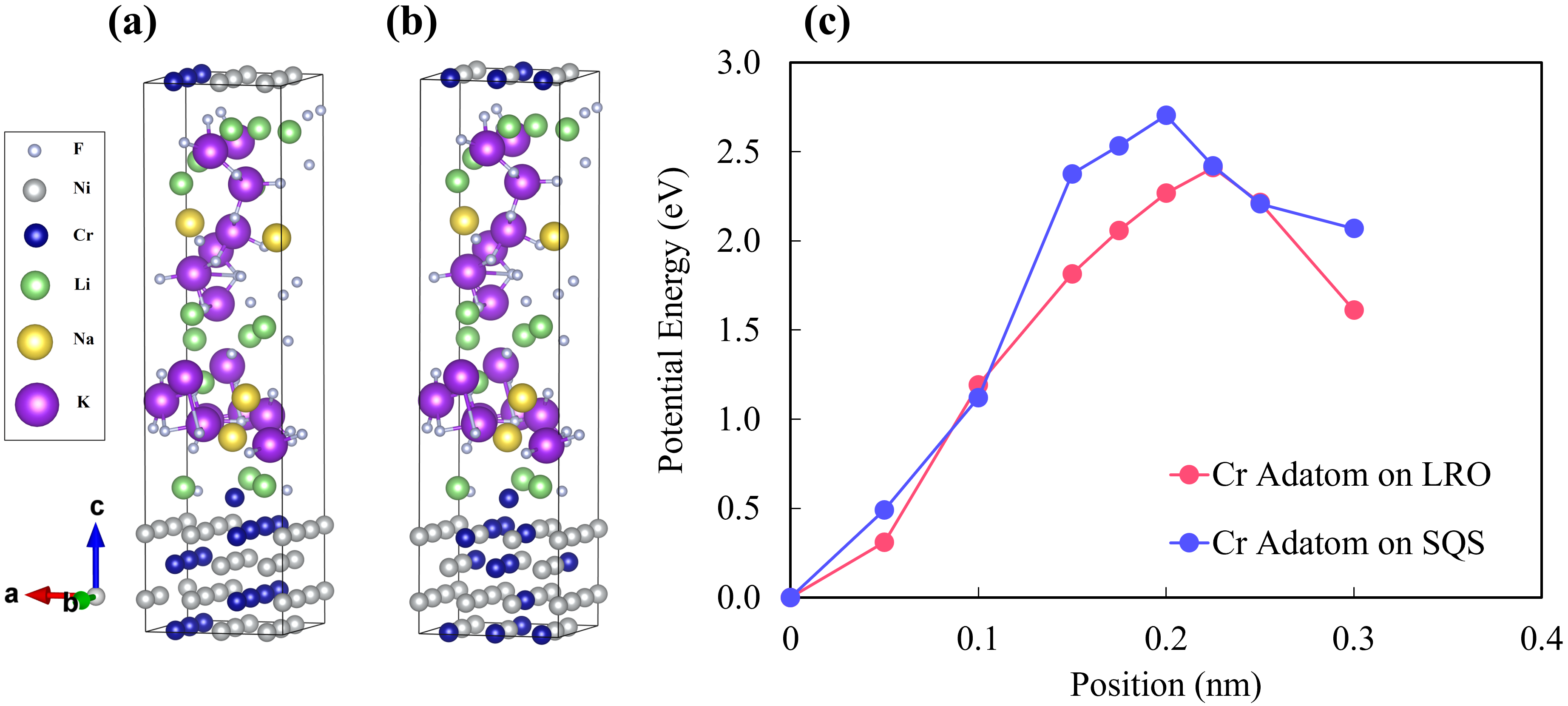}
	\caption{Cr adatom dissolution from Ni–33at.\%Cr alloys in molten FLiNaK: (a) LRO slab, (b) SQS slab, and (c) potential energy profiles showing a lower dissolution barrier for LRO compared to SQS.}
	\label{fig:dft}
\end{figure}

Finally, it is important to note that Cr dissolution into molten salt depends not only on diffusion but also on the dissolution barrier from the alloy surface. To assess whether dissolution energetics differ between LRO and RSS, we performed DFT calculations to evaluate the Cr dissolution barrier, which captures bond breaking and charge transfer more realistically than MD. Following the stepwise Cr detachment approach in \cite{arkoub2024first}, we calculated the potential energy profile of a Cr adatom interacting with equilibrated FLiNaK salt for both an LRO Ni\textsubscript{2}Cr (110) slab and an SQS slab (Figure \ref{fig:dft}a,b). The computed energy profiles (Figure \ref{fig:dft}c) reveal a slightly lower energetic barrier for Cr dissolution in the LRO alloy (2.4 eV) compared to the SQS model (2.7 eV). Both values are substantially reduced relative to the dissolution of Cr adatom from Ni(100) in vacuum (4.0 eV) \cite{arkoub2024first}, suggesting the influence of the molten salt environment and consistent with the solvent-mediated dissolution mechanism reported by Ke et al. \cite{ke2020dft}. The reduced barrier originates from weakened Cr–Ni bonding from F–Cr electronic interactions at the interface, which lowers the detachment energy. These results demonstrate that LRO accelerates corrosion through a dual mechanism: percolating Cr networks that promote diffusion and a slightly lowered energetic barrier that promotes interfacial dissolution.

\section{Conclusion}

This work provides the first atomistic evidence that chemical ordering fundamentally alters early-stage corrosion kinetics in Ni-Cr alloys. Using Ni–33at.\%Cr exposed to molten FLiNaK, we show that LRO alloys experience accelerated Cr dissolution, deeper surface degradation, and enhanced near-surface diffusion compared to SRO and RSS configurations. The accelerated corrosion in LRO alloys arises from the interplay of microstructural features and corrosion dynamics. Selective Cr leaching destabilizes surface layers and triggers rapid Ni–Cr counter-diffusion, while the ordered Ni\textsubscript{2}Cr lattice forms continuous Cr networks that serve as percolation pathways feeding Cr to the interface. This connectivity acts as a dissolution highway in fluoride salts. First-principles calculations further reveal that LRO lowers the energetic barrier for Cr detachment relative to the disordered alloy. Together, these results highlight that the topology of atomic ordering affects corrosion resistance. Disrupting Cr network connectivity through controlled SRO or alloying (e.g., with Mo) may suppress percolation-driven corrosion.

\section*{Acknowledgments}

This work was funded by the Laboratory Directed Research and Development (LDRD) Program at Idaho National Laboratory under grant number 23A1070-147FP. Part of the research was supported by the Department of Energy through the Nuclear Science User Facilities award NSUF 24-5018. Additional support was provided by the National Science Foundation (NSF) through CAREER Award No. 2340019. This research made use of Idaho National Laboratory’s High Performance Computing systems located at the Collaborative Computing Center and supported by the Office of Nuclear Energy of the U.S. Department of Energy and the Nuclear Science User Facilities under Contract No. DE-AC07-05ID14517.

\bibliographystyle{elsarticle-num}
\bibliography{references}



\end{document}